\documentclass[twocolumn,prd,nofootinbib,aps,prl,floats,floatfix,amsmath,amssymb,longbibliography,preprintnumbers,secnumarabic]{revtex4-1} %
\usepackage[english]{babel}

\usepackage{comment}

\usepackage[letterpaper,top=2cm,bottom=2cm,left=3cm,right=3cm,marginparwidth=1.75cm]{geometry}

\usepackage{amsmath}
\usepackage{graphicx}
 \usepackage{xcolor}
\usepackage[colorlinks=true,citecolor=red]{hyperref}

\newcommand{\be}{\begin{equation}}
\newcommand{\ee}{\end{equation}}
\def\sfrac#1#2{{\textstyle{#1\over #2}}}
\newcommand{\bea}{\begin{eqnarray}}
\newcommand{\eea}{\end{eqnarray}}
\newcommand{\nn}{\nonumber}

\begin{document}
\preprint{IPMU26-0008}

\title{Pathologies of dimension-zero scalar fields}
\author{James M.\ Cline}
\email{jcline@physics.mcgill.ca}
\affiliation{McGill University Department of Physics \& Trottier Space Institute, 3600 Rue University, Montr\'eal, QC, H3A 2T8, Canada}
\author{Anamaria Hell}
\email{anamaria.hell@ipmu.jp}
\affiliation{Kavli IPMU (WPI), UTIAS,
The University of Tokyo,
Kashiwa, Chiba 277-8583, Japan}
\affiliation{Center for Data-Driven Discovery, Kavli IPMU (WPI), UTIAS,
The University of Tokyo, Kashiwa, Chiba 277-8583, Japan}

\begin{abstract}
It has been claimed in a series of papers that scalar fields
with a fourth-order Lagrangian $\sim(\Box\varphi)^2$ can solve the cosmological constant problem by canceling the loop contributions from standard model fields, and that their fluctuations can be the source of the primordial density perturbations of the Universe, without the need for inflation.  We dispute these claims. The spectrum of the theory includes a ghost, which leads to classical instabilities and quantum violation of unitarity.   We
show that the new scalar particles cannot cancel the standard model contributions to the cosmological constant, unless they include a unitarity-violating ghost at the quantum level.  Further, the coupling of such scalars to the particles of the standard model induces a confining fifth force which rules it out as a source of density perturbations in the early Universe.

\end{abstract}

\maketitle

\section{Introduction}
In our quantum field theory courses, we have been taught that theories
whose Lagrangians have more than two powers of derivatives are to be avoided, except in the case of effective field theories where expansions in powers of derivatives are restricted to be small corrections.  Allowing such terms to correspond to arbitrarily high momenta results in additional poles in the propagator whose residues have the wrong sign: ghosts.  Such degrees of freedom lead to negative probabilities, hence loss of unitarity, or negative energies, depending upon boundary conditions imposed on the fields \cite{Cline:2003gs}.

Nevertheless, a particular Weyl-invariant fourth-order action for scalar fields coupled to gravity has received some attention in the literature,
\begin{equation}\label{D4sc}
 S=-\int d^4x\sqrt{-g}\varphi\Delta_4\varphi\,,
\end{equation}
where
\begin{equation} \label{Delta4}
\Delta_4=\Box^2+2R^{\mu\nu}\nabla_{\mu}\nabla_{\nu}-\frac{2}{3}R\Box+\frac{1}{3}(\nabla^{\mu}R)\nabla_{\mu}\,,
\end{equation}
being originally motivated by
studies of conformal supergravity \cite{Kaku:1978nz}.\footnote{This is claimed by Ref.\ \cite{Boyle:2021jaz} to be the unique such operator; however Ref.\ \cite{Fradkin:1982xc} notes that it can be supplemented with a contribution from the square of the Weyl tensor, with an arbitrary coefficient.}   The action (\ref{D4sc}) after integration by parts is a fourth-order theory containing squared curvatures, ${\cal L}\ni R_{\mu\nu}^2-R^2$.  As noted in Ref.\ \cite{Fradkin:1981jc}, it gives rise to 
a propagating scalar graviton mode $\varphi\sim h_\mu^\mu$ whose action
corresponds to Eq.\ (\ref{D4sc}).  The authors of Refs.\ \cite{Fradkin:1981jc,Fradkin:1982xc} were aware of the problem of ghosts in conformal supergravity, hoping that some solution would eventually be found, while investigating its  interesting properties.

As noted in Ref.\ \cite{Boyle:2021jaz}, the minus sign in
Eq.\ (\ref{D4sc}) applies in Minkowskian signature $(-+++)$,
and becomes $+$ when Wick-rotating to Euclidean signature.  It will be clear from the results we derive that the sicknesses of this theory cannot be cured by choosing the opposite sign for the overall action, since the fourth order Lagrangian typically describes both a ghost and a nonghost degree of freedom.

A scalar with the action (\ref{D4sc}) was used in Ref.\ \cite{Riegert:1984kt} as an
auxiliary field which when integrated out yields a nonlocal action  
whose variation gives the trace anomaly.
This reference also included linear couplings between $\varphi$ and curvature invariants that led to effects already at tree-level.
There was no suggestion there that $\varphi$ should be treated as a physical field.   Unlike a conventional scalar field, it has mass dimension zero.

The present work is prompted by a more recent series of papers in which it was proposed that adding 36 of such dimension-zero scalars
to the standard model of particle physics (minus the Higgs field) will solve the cosmological constant problem \cite{Boyle:2021jaz,Turok:2023amx,Boyle:2025bxf}, as well as provide the primordial density perturbations of the Universe.
They claimed that a gauge symmetry of the Lagrangian circumvents its ghost problem.
Here we will give a correct analysis of the theory, in both a flat background and in the Friedmann-Robertson-Walker metric, and we will elucidate some of the errors made in Refs.\ \cite{Boyle:2021jaz,Turok:2023amx,Boyle:2025bxf}, that invalidate their claims.

\section{The ghost problem}
The easiest way to see the presence of a ghost in the higher-derivative theory is to rewrite it as a lower-derivative Lagrangian using an auxiliary field.  For later convenience, we first formulate it in a general gravitational background.  
Using $\nabla^{\mu}R_{\mu\nu}=\frac{1}{2}\nabla_{\nu} R$ and integration by parts, the higher-derivative scalar action is equal to
\begin{equation}
    S=\int d^{\,4}x\sqrt{-g}\Bigl[\frac12(\Box\varphi)^2
-R^{\mu\nu}\partial_\mu\varphi\partial_\nu\varphi
+\frac13 R (\partial \varphi)^2 \Bigr].
\end{equation}
One can show that it is invariant under Weyl transformations $g_{\mu\nu}(x)\mapsto \Omega^2(x)g_{\mu\nu}(x)$,
$\varphi(x)\mapsto \varphi(x)$.
Rewriting the first term using a Hubbard-Stratonovich field $\sigma$, the action becomes
\twocolumngrid
\bea
S&\simeq& \int d^{\,4}x\sqrt{-g}\Bigl[
-\partial_\mu \sigma \partial^\mu\varphi
-R^{\mu\nu}\partial_\mu\varphi\partial_\nu\varphi\nn\\
&+&\sfrac13 R (\partial \varphi)^2
-\sfrac12\sigma^2
\Bigr]
\nonumber\\ &=& \!\!\int d^{\,4}x\sqrt{-g}\Bigl[
-\sfrac12
\begin{pmatrix} \partial_\mu \varphi \\\partial_\mu \sigma\end{pmatrix}^{\!\!T}\!\!\!\!
\begin{pmatrix}
R^{\mu\nu}\!-\!\frac{1}{3}Rg^{\mu\nu} & g^{\mu\nu}
\\ g^{\mu\nu} & 0
\end{pmatrix}\nn\\
&\times&
\begin{pmatrix} \partial_\mu \varphi \\ \partial_\mu \sigma\end{pmatrix}-\sfrac12\sigma^2\Bigr]. \label{Stwo-derivative}
\eea
\twocolumngrid
If the gravitational background is weak, one can ignore $R^{\mu\nu}-\sfrac13 R g^{\mu\nu}$ compared to $g^{\mu\nu}$ in Eq.\ (\ref{Stwo-derivative}). 
Then since the eigenvalues of the kinetic mixing matrix $\left({0\atop 1}\,{1\atop 0}\right)$ are $\pm 1$, there is clearly a ghost in the spectrum.  Even if the gravitational field is strong, the determinant of the mixing matrix remains unchanged, so the ghost cannot be circumvented in this case either.

One can observe already at the classical level that the system 
is subject to an instability, owing to the negative energy carried by the ghost, since the Hamiltonian density (in Minkowski space)
\be
    {\cal H} = \dot\varphi\,\dot\sigma +\vec\nabla\varphi\cdot\vec\nabla\sigma + \sfrac12\sigma^2
    \label{Hameq}
\ee
is unbounded from below.  This is again a consequence of the negative determinant of the kinetic matrix.  The equations of motion are
\be
\label{EOMS}
    \ddot\varphi -\nabla^2\varphi = -\sigma,\quad \ddot\sigma -\nabla^2\sigma= 0,
\ee
which have the homogeneous solution 
\be
    \sigma = \sigma_0 + \omega t,\quad \dot\varphi = \dot\varphi_0
    -\sigma_0 t-\sfrac12 \omega t^2\,.
\ee
The corresponding Hamiltonian is
\be
    {\cal H} = \omega\dot\varphi_0 + \sfrac12\sigma_0^2
\ee
which can take arbitrarily negative values.

Moreover, inhomogeneous configurations may fluctuate 
with arbitrarily negative local values of ${\cal H}$, as illustrated by the
wave-like solution
\be
    \varphi = A\, p\!\cdot\! x\, \sin k\!\cdot\! x,\quad
        \sigma = -2A\,p\!\cdot\! k\,\cos k\!\cdot\! x,
\ee
where $k^\mu$ is a null vector ($k^2=0$) and $p$ is any 4-vector such 
that $p\!\cdot\! k\neq 0$.    Then there is an oscillating contribution to the energy density that is linear in $p\!\cdot\! x$, 
\be
    {\cal H} \sim 4 A^2 \omega^2\, p\!\cdot\! k\, p\!\cdot\! x\,
        \cos k\!\cdot\! x\,\sin k\!\cdot\! x
\ee
whose amplitude is unbounded as $|x|\to\infty$.

In this noninteracting theory, the sicknesses inherent to ghosts are ameliorated, since the energy is conserved.  As soon as the negative energies can be offset by positive energies of normal particles interacting with $\varphi$, further instabilities become apparent, as we demonstrate in the next section, by studying the dynamics in cosmological backgrounds.

In Ref.\ \cite{Boyle:2021jaz}, it was argued that there are actually no propagating physical states associated with the field $\varphi$ once the ``gauge symmetry'' is taken into account,
where $\varphi\to \varphi+\psi$ for harmonic functions $\Box\psi =0$.\footnote{The authors of \cite{Boyle:2021jaz} then seem to forget about this argument when they subsequently claim that $\varphi$ has scale-invariant fluctuations in the early Universe, that can replace inflationary ones.}  This argument comes from section 10.2 of the textbook \cite{Bogolyubov:1990kw}, but we believe that exposition was  misinterpreted by Ref.\ \cite{Boyle:2021jaz}.  A full discussion of this issue will be given in Section \ref{nogauge}, where we demonstrate that no dynamical degrees of freedom are eliminated by this symmetry, unless additional assumptions are invoked.

\section{FLRW background}
\label{FLRW}

In curved spacetime, the characters of the degrees of freedom can be different than in Minkowski space.  Nonvanishing background values for $\varphi$ and the curvature $R$ causes mixing between $\varphi$ and the scalar
components of the metric perturbations.  This can produce
qualitatively different properties for the fluctuations.

We consider a Friedmann-Lema\^itre-Robertson-Walker form for the metric, including a lapse function $N$, and a spatially homogeneous background value for $\varphi$,
\bea\label{bcgIII}
   ds^2&=&-N^2dt^2+a(t)^2\delta_{ij}dx^idx^j\,,\nn\\\varphi &=& \varphi(t)\,. 
\eea
By substituting this ansatz into the action, varying it with respect to the lapse, and then setting $N=1$, we find the constraint equation
\begin{equation}\label{constraint}
2 \dot{\varphi} \dddot{\varphi}-{\ddot{\varphi}}^{2}+\frac{4 \dot{\varphi} \dot{a} \ddot{\varphi}}{a}+\frac{2 \dot{\varphi}^{2} \ddot{a}}{a}+\frac{\dot{\varphi}^{2} \dot{a}^{2}}{a^{2}}
=0\,.
\end{equation}
As a consequence of Weyl invariance,  varying with respect to the scale factor gives
the same equation. Variation with respect $\varphi$ gives its equation of motion
\bea
\label{phieom}
    2 a^{3} \ddddot{\varphi}&+&\left(12 \dot{a} \dddot{\varphi}+2 \dot{\varphi} \dddot{a}+8 \ddot{a} \ddot{\varphi}\right) a^{2}\nn\\
    &+&\left(14 \dot{a}^{2} \ddot{\varphi}+8 \dot{\varphi} \ddot{a} \dot{a}\right) a+2 \dot{\varphi} \dot{a}^{3}
=0\,.
\eea
We can regard Eqs.\ (\ref{constraint},\ref{phieom}) as determining the higher derivatives $\{\ddddot{\varphi}(t),\dddot{\varphi}(t)\}$ of $\varphi$ in terms of the remaining
quantities.

We found three simple analytic solutions to the background equations, which will be useful for examples given below.
For a matter dominated universe with $a(t) \sim t^{2/3}$,
$\varphi(t) \sim t$ is an exact solution.  For a de Sitter background with $a(t) \sim e^{Ht}$, there are two exact solutions, with $\varphi\sim 1/a$ and $\varphi\sim 1/a^3$, respectively.  These cannot be superposed, even though the equation of motion for $\varphi$ is linear, since the constraint equation is quadratic.

We next consider the dynamical fluctuations of $\varphi$ and
components of the metric around the above background, taking
\bea
\label{dec1}
    g_{\mu\nu}&=& g_{\mu\nu}^{(0)}+\delta  g_{\mu\nu}\,,\nn\\
   \varphi&=&\varphi^{(0)}+\chi,, 
\eea
where $g_{\mu\nu}^{(0)}$ and $\varphi^{(0)}$ are the background fields. The metric perturbations can be decomposed according to their transformations under spatial rotations:
\bea
\label{dec2}
        \delta g_{00}&=& -2\phi\nn\\
        \label{vectorfluct}
        \delta g_{0i}&=&a(t)\left(S_i+B_{,i}\right)\\
        \delta g_{ij}&=&a^2(t)\left(2\psi \delta_{ij}+2E_{,ij}+F_{i,j}+F_{j,i}+h_{ij}^T\right)\nn
\eea
where the vector perturbations are divergenceless, 
$S_{i,i}=0$ and $F_{i,i}^T=0$, 
while tensor perturbations divergenceless and traceless, 
$h_{ij,j}^T=0$ and $h_{ii}^T=0$.
At the leading order, the metric, vector and tensor perturbations decouple from each other, allowing us to study them separately. 

\begin{figure*}[t]
\centerline{\includegraphics[width=\columnwidth]{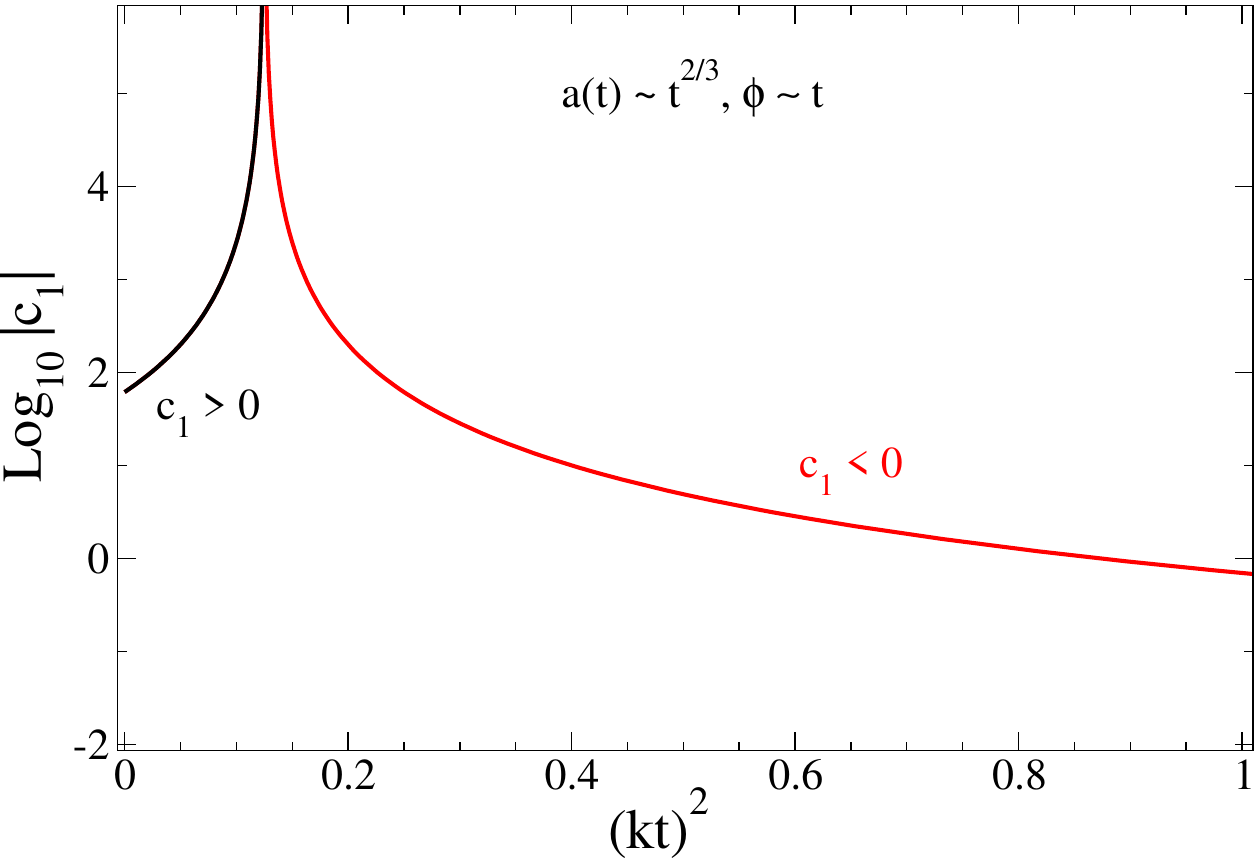}
\includegraphics[width=\columnwidth]{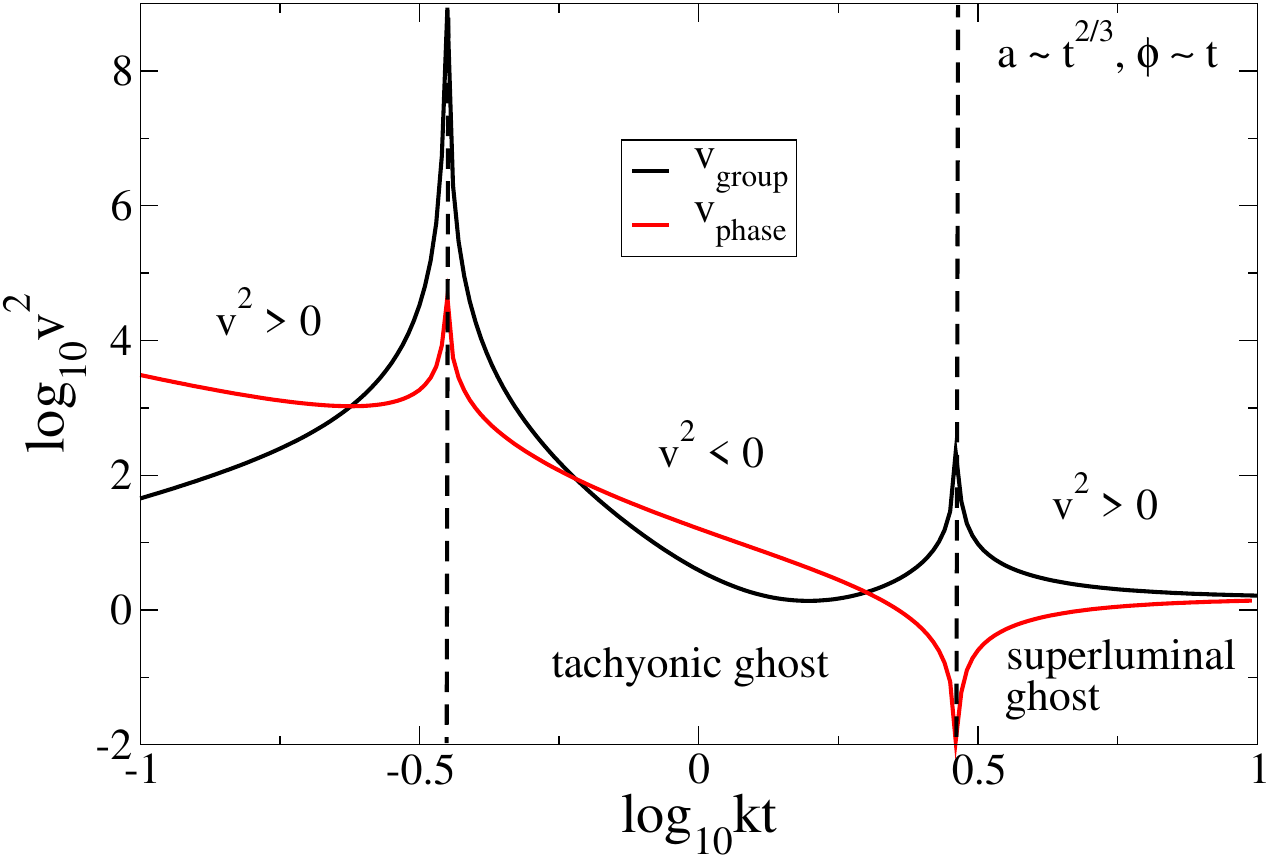}}
\caption{Left: $c_1$ coefficient for scalar perturbations, Eq.\ (\ref{scalar-pert}), versus dimensionless wave number squared, $\hat k^2 = k^2 t^2$, in the cosmological background $a(t)\sim t^{2/3}$, $\phi(t)\sim t$.  
For small $\hat k$, it is negative, indicating that the perturbations are ghosts.  Right: squared group and phase velocities of the perturbations in this case. }
\label{fig:c1plot}
\end{figure*}

\begin{figure}[t]
\centerline{\includegraphics[width=\columnwidth]{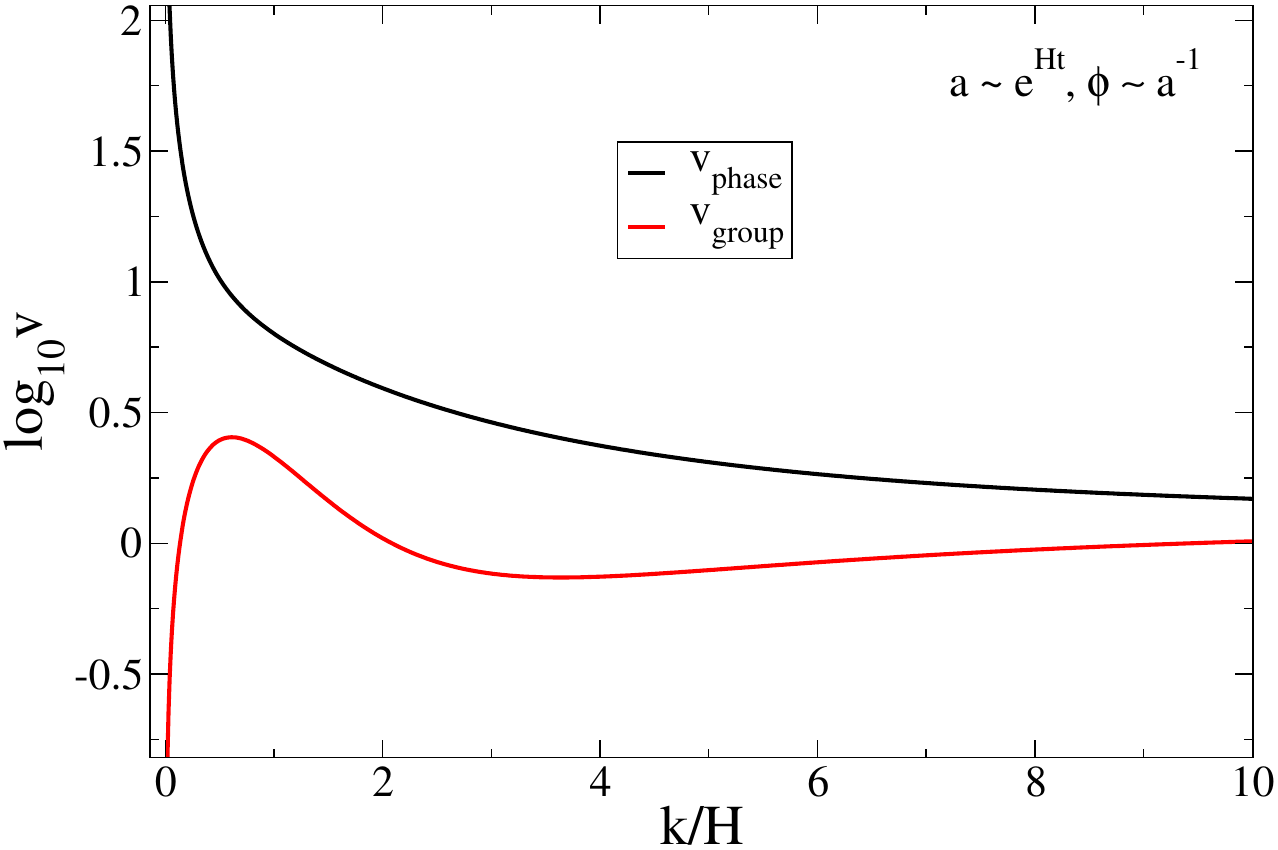}}
\caption{Magnitude of phase and group velocities for scalar ghost perturbations around the inflationary background with $a\sim e^{Ht}$ and $\varphi\sim 1/a$.}
\label{fig:vplot}
\end{figure}

\medskip

\subsection{Scalar fluctuations around the background}

Adopting the Newtonian gauge with $E=0$ and $B=0$,  we expand the action to the second order in the scalar perturbations. After integrations by parts, this gives the Lagrangian density
\begin{equation}
    \mathcal{L}_S=-a^3\ddot{\chi}^2-2a^3\dot{\varphi}\ddot{\chi}(\dot{\psi}-\dot{\phi})+ \dots
\end{equation}
where $\dots$ denotes the remaining terms, which contain only first or zeroth derivatives of the fields (for the full expression see the Appendix).  For brevity, we drop the $^{(0)}$ superscript of the background field.
The second-order derivatives of $\chi$ are coupled to the gravitational potentials, and this interaction is multiplied $\dot\phi$. Therefore, when $\dot\phi\ne 0$, the theory may
have qualitatively different behavior than the case $\dot{\varphi}=0$, as we shall demonstrate.

In order to further explore its content, we substitute
\begin{equation}
    \psi=\psi'+\phi. 
\end{equation}
This makes the scalar metric perturbation $\phi$ entirely decouple from  the action, including the $\dots$ terms (see appendix for details): 
\begin{equation}\label{2ndorder}
\mathcal{L}_S=-a^3\ddot{\chi}^2-2a^3\dot{\varphi}\ddot{\chi}\dot{\psi}'-a^3\dot{\varphi}^2\dot{\psi'}^2+\dots
\end{equation}

We further introduce an auxiliary scalar $\sigma$
to rewrite the higher-derivative terms as 
\begin{equation}
\mathcal{\Tilde{L}}_S=a^3\sigma^2-2a^3\sigma(\ddot{\chi}+\dot{\varphi}\dot{\psi}')+\dots
\end{equation}
where $\sigma$ obeys the equation of constraint
$\sigma=\ddot{\chi}+\dot{\psi}'\dot{\varphi}$.
 By partially integrating and substituting 
 \begin{equation}
    \sigma = \sigma' -\left(2 \ddot{\varphi} a+\dot{a} \dot{\varphi}\right) \psi' \frac{1}{a }
\end{equation}
the kinetic term for $\psi'$ is eliminated, and it becomes
another auxiliary field whose equation of motion is a constraint.  By integrating out $\psi'$, the action becomes
a function of only $\sigma'$ and $\chi$.  
{By  expressing the fields in the momentum space as}
\begin{equation}\label{momentumsp}
    X=\int \frac{d^{\,3}k}{(2\pi)^{3/2}}{X}_{\vec{k}}\,e^{i\vec{k}\vec{x}},
\end{equation}
where $X$ stands for either $\sigma'$ or $\chi$, and substituting 
\begin{equation}
    \begin{split}
        \sigma_{\vec{k}}' &=\tilde\sigma_{\vec{k}}\\&-\frac{1}{6a^2\dot{\varphi}^2} \left(2  \dot{\varphi}^{2}k^2-3 a^{2} {\ddot{\varphi}}^{2}+3 \dot{\varphi}^{2} \dot{a}^{2}\right) \chi_{\vec{k}},
    \end{split}
\end{equation}
the kinetic term for $\chi$ is also removed, making it too an auxiliary field. 
Integrating out the $\chi$ field, the final form of the scalar perturbation Lagrangian has only a single dynamical scalar fluctuation $\tilde\sigma$, 
\bea
\label{scalar-pert}
\mathcal{\Tilde{L}}_{Sk}&=&c_1(t,k^2)\,\dot{\tilde\sigma}_{\vec{k}}\,\dot{\tilde\sigma}_{-\vec{k}}\nn\\
&-&c_2(t,k^2)\,k^2\,\tilde\sigma_{-\vec{k}}\,\tilde\sigma_{\vec{k}},
\eea
where $c_{1,2}(t,k^2)$ are complicated functions of the background fields $\varphi(t,\vec x)$ and $a(t)$, and the momentum of the perturbations, whose full expressions are given in the Appendix.   They simplify in the limit $k\to\infty$, and lead to a superluminal propagation speed 
\be
    c_s = \sqrt{a^2 c_2 \over c_1} = \sqrt{3\over 2}\,,
\ee
where the factor of $a^2$ takes into account the redshifting of the physical momentum.

For finite $k$, the phase and group velocities differ from each other, and their behavior depends upon the background $a(t)$ and $\varphi(t)$ solutions.  As an example, consider the matter-dominated background $a(t)\sim t^{2/3}$, where $\varphi(t) = \dot\varphi_0\, t$.\footnote{Here we assume that $\varphi$ is a spectator field, not dominating the energy density of the Universe, hence its equation of state need not be that of nonrelativistic matter.}.  The propagation speed can be expressed as a function of a dimensionless wave number $kt = 3k/2H$, where $t$ is cosmic time and $H$ is the Hubble parameter.  For this background, $c_1$ is negative at large $k$, indicating the perturbations are (superluminal) ghosts.
For intermediate $k$, there is a region where both $c_1$ and $c_2$ are negative, resulting in tachyonic ghosts.   At small $k$, these pathologies are avoided,
but the phase velocity diverges as $k\to 0$, while the group velocity goes to zero.

For an inflationary background with $a\sim e^{Ht}$ and $\varphi\sim 1/a$, the scalar perturbations are ghosts at all wavenumbers,  and like the matter-dominated case, they become superluminal at wavenumbers $k/H\gtrsim 10$, reaching the same
asymptotic speed of $\sqrt{3/2}$ as in that case.  The phase and group velocities diverge or approach zero respectively as $k\to0$, as shown in Fig.\ \ref{fig:vplot}.  The other inflationary solution with $\varphi\sim 1/a^3$ has similar behavior.  

\subsection{Vector fluctuations}
It turns out that there are no physical vector fluctuations in the theory, even though $S_i$ is {\it a priori} present in the metric decomposition (\ref{vectorfluct}).  By expanding its action to second order and integrating several times by parts, we its Lagrangian density 
\begin{equation}
   \mathcal{L}_V= \frac{2}{3} \dot{\varphi}^{2} a k^{2} S_i\Delta S_i 
\end{equation}
in the Poisson gauge with $F_i=0$, which corresponds to the gauge invariant variable $V_i=S_i-\dot{S}_i$. Clearly, $S_i$ does not propagate since there are no time derivatives. It satisfies the constraint $\Delta S_i =0$, and can therefore be set to zero.

\subsection{Tensor fluctuations}
For simplicity, we have not included the $\sqrt{-g}R$ term of the gravitational action in this section; see Section \ref{incgrav}.  It contributes a large positive kinetic term to the tensor fluctuations, which would be negative without it (see appendix).  Hence we do not further consider them here.

\subsection{Limit of Minkowski space}
If one takes the limit $a(t)\to $ constant, the theory should revert to that of the Minkowski background; however it is possible to still have a nontrivial $\varphi(t)$ field.  The equation of constraint (\ref{constraint})
reduces to $2 \dot{\varphi} \dddot{\varphi}={\ddot{\varphi}}^{2}$, while the equation of motion is $\ddddot\varphi = 0$.  Supposing for simplicity a power law solution $\varphi\sim t^p$, the only nontrivial solution is with $p=1$.  In this case, the scalar fluctuation is a ghost with propagation speed 
$\sqrt{3/2}$, independently of $k$, and the tensor fluctuation is a tachyonic ghost, as in the previous subsection.  A strong-coupling obstruction 
is expected in the limit $\dot\phi\to 0$ as discussed there. 

\section{Inclusion of dynamical gravity}
\label{incgrav}
In the foregoing section, we neglected the usual graviton kinetic term
$\sqrt{-g}R$ from the Einstein-Hilbert action, which can mix with the kinetic operators considered above.  This neglect is expected to be a good approximation when the background $\dot\phi\ll M_P$ (the reduced Planck mass), since the new fluctuations become more strongly coupled than gravity in that regime.

Here we relax that restriction and include the $\sqrt{-g}R$ contribution, considering the action 
\begin{equation}\label{RD4}
 \begin{split}
     S_{g\Delta_4}&=\int d^4x\sqrt{-g}\left(\frac{1}{2}M_P^2 R- \Box\varphi\Box\varphi\right.\\&\left.+2R^{\mu\nu}\nabla_{\mu}\varphi\nabla_{\nu}\varphi-\frac{2}{3} R \nabla_{\mu}\varphi\nabla^{\mu}\varphi\right).
 \end{split}
\end{equation}
We will consider three cosmological regimes: (1) $\varphi$ has a background solution that dominates the energy density of the Universe; (2) $\varphi$ as well as other sources of stress-energy contribute significantly to the expansion rate; and (3)
is a spectator field.

The action (\ref{RD4}) is not the full higher-dimensional gravitational action considered in \cite{Boyle:2025bxf}, but contains only the leading term, the Ricci scalar. We have omitted the square of the Weyl tensor as well as the derivatives of the Ricci scalar of the form $\Box R$. Including them would  remove the constraint from the gravitational potential $\phi$ which otherwise holds (see below Eq.\ (\ref{Rsubs4})), 
exacerbating the ghost problem in the scalar sector.
Including the square of the Weyl tensor introduces a ghost mode for the tensor modes, as well as two vector modes\footnote{See however \cite{Maldacena:2011mk, Hell:2023rbf}, for the potential resolution of the gravitational ghost via boundary conditions. } \cite{Stelle:1977ry}. In the following therefore, we take the Lagrangian to be Eq.\ (\ref{RD4}).

\subsection{Scalar field with non-vanishing background}
The inclusion of the Ricci scalar breaks the conformal invariance that was present in the original $\Delta_4$ theory. Assuming the background (\ref{bcgIII}), varying the action with respect to the lapse $N$, and then setting $N=1$, we find the constraint equation
\begin{equation}\label{constraintR}
  \begin{split}
        &\ 3 M_P^{2}H^2 +\dot{\varphi} \dddot{\varphi}+{\ddot{\varphi}}^{2}-\frac{2 \ddot{a} \dot{\varphi}^{2}}{a}\\&-{4 H \dot{\varphi} \ddot{\varphi}}-H^2 \dot{\varphi}^{2}
=0\,
  \end{split}
\end{equation}%\label{accR}
\noindent where $H=\dot a/a$.  Variation with respect to the scale factor  gives
\begin{equation}
 \begin{split}
     &-{\ddot{\varphi}}^{2}+2 \dot{\varphi} \dddot{\varphi}+3 M_P H^{2}+H^2 \dot{\varphi}^{2}\\&+ (6 M_P+ 2\dot\phi^2) {\ddot{a}\over a}+4 H \dot{\varphi} \ddot{\varphi}=0,
 \end{split}
 \label{scale-factor-eq}
\end{equation}
while the equation of motion for the scalar field is given by:
\begin{equation}
 \begin{split}
        &2 \ddddot{\varphi} +12 \dddot{\varphi} H+2 \dot{\varphi} {\dddot{a}\over a}+8 \ddot{\varphi} {\ddot{a}\over a}\\& +14 \ddot{\varphi} H^2+8 H {\ddot{a}\over a} \dot{\varphi}+2 \dot{a}^{3} \dot{\varphi}
=0\,.
 \end{split}
 \label{scalar-eom}
\end{equation}
Eqs.\ (\ref{scale-factor-eq},\ref{scalar-eom}) can be combined to show that 
\begin{equation}\label{scaleFR}
    {\ddot{a}\over a} = -H^2\,,
\end{equation}
which describes a radiation-dominated Universe.  Since $\varphi$ is scale-invariant, this result could be anticipated, and we find 
 the solutions
\bea
\label{bkg-solns}
        a(t)&=&_0\sqrt{t}\\
        \varphi(t)&=& 
c_{1} +c_{3}\left(M_P^{2}+{c_{2}^{2}\over3}\right) \sqrt{t} +c_{2} t + {t^{{3}/{2}}\over c_{3}},\nn
\eea
where $a_0, c_1, c_2$ and $c_3$ are integration constants.

To find the action for the scalar modes, we decompose the metric and the scalar field according to (\ref{dec1}) and (\ref{dec2}), and expand the action up to second order in the perturbations, while fixing the longitudinal gauge. After performing several integrations by parts, and substituting 
\begin{equation}\label{Rsubs1}
    \psi=\psi'+\phi,
\end{equation}
$\phi$ does not decouple from the remaining fields, as it did in the case of the pure $\Delta_4$ theory, and higher derivatives of fields still remain. To reduce their order, we introduce the auxiliary field $\sigma$ as before, which now satisfies the constraint
\begin{equation}\label{Rsubs2}
    \sigma= \ddot{\chi} + \dot{\varphi}\dot{\psi}'\,
\end{equation}
obtained by varying the modified action with respect to it. After several integrations by parts, and by substituting
\begin{equation}\label{Rsubs3}
    \sigma=\sigma' - (\dot{\varphi}\dot{a} + 2\ddot{\varphi}a)\frac{\psi'}{a},
\end{equation}
followed by 
\begin{equation}\label{Rsubs4}
    \psi'=\psi''-\phi, 
\end{equation}
 we find that $\phi$ becomes a nonpropagating field. 
 Its equation of motion gives a new constraint, which upon substitution back into the action renders the equations of motion second order, involving only 
 $(\psi'', \chi, \sigma')$. Fourier transforming the fields using the convention (\ref{momentumsp}), and introducing further substitutions
\begin{equation}\label{Rsubs5}
    \psi_{\vec{k}}''=\tilde{\psi}_{\vec{k}} + \frac{\dot{a}}{\dot{a}\dot{\varphi}}\chi_{\vec{k}},
\end{equation}
and
\begin{equation}\label{Rsubs6}
   \begin{split}
        \sigma_{\vec{k}}'&=\sigma_{\vec{k}}'' + \frac{1}{6a^2\dot{\varphi}^2}\left(9M_p^2\dot{a}^2\right.\\&\left. - 3\dot{a}^2\dot{\varphi}^2 + 3\ddot{\varphi}^2a^2 + 2\dot{\varphi}^2k^2\right)\chi_{\vec{k}}, 
   \end{split}
\end{equation}
causes $\chi$ to also become nonpropagating.
Using its constraint, the action becomes a function of only the two fields $\sigma''$ and $\tilde{\psi}$.

To diagonalize the kinetic terms, we substitute
\begin{equation}
    \sigma_{\vec{k}}''=\tilde{\sigma}_{\vec{k}}-\frac{3 \left(3 M_P^{2} \dot{a}-9 \dot{\varphi}^{2} \dot{a}-2 \ddot{\varphi} \dot{\varphi} a\right) M_P^{2}}{8 a \dot{\varphi}^{3}}\tilde{\psi}_{\vec{k}}
\end{equation}
and find the  Lagrangian density in momentum space, 
\begin{equation}
    \begin{split}  \mathcal{\tilde{L}}_{Sk}&=d_1(t,k)\dot{\tilde{\psi}}_{\vec{k}}\dot{\tilde{\psi}}_{-\vec{k}} +d_2(t,k)\dot{\tilde{\sigma}}_{\vec{k}}\dot{\tilde{\sigma}}_{-\vec{k}}\\&+\frac{1}{2}d_3(t,k)\left(\dot{\tilde{\psi}}_{\vec{k}}\dot{\tilde{\sigma}}_{-\vec{k}}+\dot{\tilde{\sigma}}_{\vec{k}}\dot{\tilde{\psi}}_{-\vec{k}}\right)\\&+d_4(t,k)\tilde{\psi}_{\vec{k}}\tilde{\psi}_{-\vec{k}}+d_5(t,k)\tilde{\sigma}_{\vec{k}}\tilde{\sigma}_{-\vec{k}}\\&+\frac{1}{2}d_6(t,k)\left(\tilde{\psi}_{\vec{k}}\tilde{\sigma}_{-\vec{k}}+\tilde{\sigma}_{\vec{k}}\tilde{\psi}_{-\vec{k}}\right), 
    \end{split}
\end{equation}
where $d_1,\dots,d_6$ are time and momentum dependent functions. Only $d_1$ and $d_2$ are needed to determine whether there are ghosts, 
\begin{equation}
    d_1(t,k)=\frac{3 \left(\frac{3}{2} M_P^{2}-4 \dot{\varphi}^{2}\right) M_P^{2} a^{3}}{4\dot{\varphi}^{2}}
\end{equation}
and
\onecolumngrid
\begin{equation}
    d_2(t,k)=-\frac{24\, a^{3} \dot{\varphi}^{4}}{54 M_P^{4}H^{2}-18 M_P^{2} \dot{\varphi}^{2}H^{2}-36 M_P^{2} \dot{\varphi} \ddot{\varphi} H-21 \dot{\varphi}^{4}H^{2}+32 \dot{\varphi}^{4} k^{2}+12 \dot{\varphi}^{3} \ddot{\varphi}H+36 {\ddot{\varphi}}^{2}\dot{\varphi}^{2}}\,.
\end{equation}
\twocolumngrid
Using the solutions for the background fields (\ref{bkg-solns}),
one can show that $d_1 < 0$ for any value of $c_3$, and $d_2<0$ in all simplifying regimes that we considered.  For example, at high momentum $k$, 
\begin{equation}
     d_2(t,k)\sim-\frac{3a^5}{4k^2}\,.
\end{equation}
Moreover, by turning on only $c_3$ in the solution (\ref{bkg-solns}),
one can show that it dominates the denominator of $d_2$ at both early and late times, leading to $d_2 < 0$.
Since  like $d_1$ it is negative, two ghosts are present
in this version of the model.

\subsection{Inclusion of matter}

In the previous subsection, only $\varphi$ was assumed to contribute to the energy density of the Universe, leading to the restriction of $a\sim t^{1/2}$ cosmological expansion.  Here we consider a toy model in which both $\varphi$ and another component, taken to be an additional scalar field $\Phi$, can contribute.  This alleviates the restriction on $a(t)$ and leads to qualitatively different behavior, as we will show.
The methodology is similar to the previous case; we here summarize the main steps.
The action is given by 
\begin{equation}
    \begin{split}
        S=S_{g\Delta_4}-\frac{1}{2}\int d^4x\sqrt{-g}\left(\partial_{\mu}\Phi\partial^{\mu}\Phi-V(\Phi)\right),
    \end{split}
\end{equation}
where the external scalar $\Phi$ EOM is
\begin{equation}
    -V_{,\Phi}-3 H \dot{\Phi} -\ddot{\Phi}=0.
\end{equation} 
Here $_{,\Phi}$ denotes the derivative with respect to $\Phi$, and $H=\dot a/a$.
Variation of the action with respect to the lapse gives the first Friedmann equation 
\begin{equation}
    \begin{split}
        &-V+H^2\left(3 M_P^{2}-\dot\varphi^2\right)-\frac{\dot{\Phi}^{2}}{2}-2 \frac{ \ddot{a}}{a}\dot{\varphi}^{2}\\& -2 \dot{\varphi} \dddot{\varphi}+{\ddot{\varphi}}^{2}- 4 H\dot{\varphi} \ddot{\varphi} =0,
    \end{split}
\end{equation}
while the variation with respect to the scale factor gives
\begin{equation}
    \begin{split}
        &-6 V -2 {\ddot{\varphi}}^{2}+4 \dot{\varphi} \dddot{\varphi}+3 \dot{\Phi}^{2}+8 \dot{\varphi} \ddot{\varphi} H\\&+2\left(3 M_P^{2} +\dot{\varphi}^{2}\right)\left(2{\ddot{a}\over a} + H^2\right)=0
    \end{split}
\end{equation}
By combining the these two equations, one can find that the equation for the scale factor which was in the absence of matter given by (\ref{scaleFR}) now generalizes to
\begin{equation}
    6 M_P^{2} \left({\ddot{a}\over a}+H^2\right)+\dot{\Phi}^{2}-4 V=0\,.
\end{equation}

To analyze the degrees of freedom, we perturb around the background values, including the external scalar field $\Phi(t)+\delta\Phi$, and assume that the background equations of motion, solved in terms of $\{\dddot{\varphi}, \ddot{a},  V_{,\Phi}(\Phi)\}$ are satisfied. The analysis is similar to the previous subsection: we follow the same steps as in Eqs.\
(\ref{Rsubs1}-\ref{Rsubs6}), 
followed by
\begin{equation}\label{chisubs}
   \begin{split}
        \chi_{\vec{k}}&=\chi_{\vec{k}}'+\frac{ \dot{\varphi}}{\dot{\Phi}}\delta\Phi_{\vec{k}}%\\&
        -\frac{4 \dot{\varphi}^{2}}{\dot{\Phi}^{2} +2 V }\sigma_{\vec{k}}''
   \end{split}
\end{equation}
which makes $\chi'$ a non-propagating field. We find its constraint, solve it, and substitute back to the action, which then becomes a function of three propagating fields, $\tilde{\psi}$, $\sigma''$ and $\Phi$. One can show that the determinant of the kinetic matrix no longer vanishes, meaning that these three fields are the degrees of freedom of the theory. 

Since the final Lagrangian takes quite a complicated form, we will omit its expression, and state the conditions under which the above modes are not ghosts. These conditions are determined by the determinant of the total kinetic matrix, determinant of the sub-matrix, corresponding to $\sigma'''$ and $\Phi$, and the kinetic term that multiplies the $\dot{\Phi}^2$ term in the Lagrangian density.\footnote{One can also choose analogously kinetic terms for $\tilde{\psi}$ and $\sigma'''$ and the corresponding sub-matrices to find an equivalent result.}  Specifically, the kinetic matrix is denoted by $\mathcal{L}\supset K_{ij}\dot{V}_{i,\vec{k}}\dot{V}_{j,-\vec{k}}$, where $V_i=(\tilde{\psi}, \sigma'',\Phi)$.  We define the (sub)determinants $K=\text{det}(K_{ij}),$ ${\rm det\,}K^1_1=\text{det}(K_{22}K_{33}-K_{23}K_{32})$, and $K^{1,2}_{1,2}=K_{33}$.  These quantities are given by
\onecolumngrid
\begin{equation}
    \begin{split}
        K=\frac{9 M_P^{4} a^{9} \dot{\varphi}^{2}}{2 \left(-3 M_P^{2}+2 \dot{\varphi}^{2}\right) V+\left(8 M_P^{2} k^{2}/a^2-6 M_P^{2} H^{2}-\dot{\Phi}^{2}\right) \dot{\varphi}^{2}+6 {\ddot{\varphi}}^{2} M_P^{2}},
    \end{split}
\end{equation}
\begin{equation}
    \begin{split}
        {\rm det\,}K^1_1=-\frac{3 \dot{\varphi}^{2} a^{6} M_P^{2}}{2 \left(-3 M_P^{2}+2 \dot{\varphi}^{2}\right) V+\left(8 M_P^{2} k^{2}/a^2-6 M_P^{2} H^{2}-\dot{\Phi}^{2} \right) \dot{\varphi}^{2}+6 {\ddot{\varphi}}^{2} M_P^{2} },
    \end{split}
\end{equation}
and 
\begin{equation}
    \begin{split}
        K^{1,2}_{1,2}=a^3\frac{\left({4 \dot{\varphi}^{2}}-6M_P^{2}\right) V+\left(6 {\ddot{\varphi}}^{2} M_P^{2}-3 \dot{\Phi}^{2} M_P^{2}-\dot{\Phi}^{2} \dot{\varphi}^{2}\right)+\left(8k^{2}/a^2-6H^2\right) \dot{\varphi}^{2} M_P^{2}}{\left( 8 \dot{\varphi}^{2}-12 M_P^{2}\right) V+\left(12 {\ddot{\varphi}}^{2} M_P^{2}-2 \dot{\Phi}^{2} \dot{\varphi}^{2}\right)+\left(16k^{2}/a^2-12 H^2\right) \dot{\varphi}^{2} M_P^{2}}
    \end{split}
\end{equation}
\twocolumngrid
Using Sylvester's criterion, the kinetic matrix is positive
definite only if all three quantities are positive:
\begin{equation}
    K >0,\quad {\rm det\,}K^1_1>0,\quad\text{and}\quad K^{1,2}_{1,2}>0\,. 
\end{equation}
However, $K$ and ${\rm det\,}K^1_1$ are proportional, with a negative ratio, since the scale factor is positive:
\begin{equation}\label{ngrel}
    K=-3M_P^2\, a^3\, {\rm det\,}K^1_1\,. 
\end{equation}
Therefore, since the scale-factor is always positive, we can see that regardless of the background, at least one of the scalar modes will always be a ghost.

We have checked that setting $\Phi=0$ gives back the results
of the previous subsection, which found two ghost modes.
By the analogous procedure, one can show that the same result holds if $\Phi$ is replaced by a $k$-essence field, 
\begin{equation}
    S=\int d^4x\sqrt{-g}\;p(X,\Phi)\,,
\end{equation}
where $X=-g^{\mu\nu}\partial_{\mu}\Phi\partial_{\nu}\Phi$.

\subsection{$\Delta_4$ scalar as a spectator}
Lastly, we consider the case when the Universe is dominated by
a perfect fluid, while treating the $\Delta_4$ scalar as a spectator field. In this case, the constraint equation becomes 
\begin{equation}
    3M_P^2H^2=\varepsilon\,,
\end{equation}
where $\varepsilon$ is the energy density, while the acceleration equation simplifies to the standard expression 
\begin{equation}
   6M_P^2\frac{\ddot{a}}{a}=-(\varepsilon+3p)\,.
\end{equation}
Therefore, depending on the external matter with equation of state $w$, with $p=w\varepsilon$, the scale factor expands
as $a\sim t^p$ with $p = 2/(3(1+w))$.
Assuming that $\varphi$ has vanishing background, 
 it decouples from the gravitational and matter perturbations at the level of quadratic action, which we will thus ignore. Then the kinetic term for $\varphi$ is given by 
\begin{equation}
   \begin{split}
        \mathcal{L}_{\chi}&=-a^3\ddot{\chi}^2+\frac{-a^3(\varepsilon+p)+4M_P^2\dot{a}^2}{2M_P^2}\dot{\chi}^2\\&-\frac{a}{2}\dot{\chi}\Delta\dot{\chi}-\frac{1}{a}\Delta\chi\Delta\chi
   \end{split}
\end{equation}
We can see that the first term is higher order in time derivatives, thus indicating that there are two scalar modes corresponding to $\chi$ -- a healthy mode, and a ghost.  

 \section{There is no gauge symmetry}
 \label{nogauge}
 The action (\ref{D4sc}), restricted to a Minkowski space background, {with vanishing background value of $\varphi$}, is invariant under transformations $\varphi\to \varphi + \alpha(x)$, for harmonic functions such that $\Box\alpha=0$, and which contribute no surface term after integration by parts.\footnote{This symmetry is broken by the curvature-dependent terms in Eq.\ (\ref{Delta4}) that do not involve $\Box$. Therefore, here we consider only the linearized theory, and neglect the higher-order contributions of the metric perturbations.}\ \   In Ref.\ \cite{Boyle:2021jaz} it was asserted that this ``gauge symmetry'' removes the dynamical degrees of freedom, thereby circumventing the problem of ghosts.  They cite an argument in the textbook \cite{Bogolyubov:1990kw} in support of their assertion.  Here we will show that this is a misunderstanding, and there is no reduction of degrees of freedom resulting from the symmetry, unless additional assumptions are made.

 Clearly the transformation $\varphi\to\varphi+\alpha(x)$ is not a gauge transformation in the usual sense, under which common physical observables remain
invariant.  Let us take the Minkowski space limit of Eq.\ (\ref{Stwo-derivative}), whose Lagrangian is $-\partial_\mu\varphi\,\partial^\mu\sigma - \frac12\sigma^2$ 
Then, for example, the stress-energy tensor 
\be
T_{\mu\nu} = \partial_\mu\varphi\,\partial_\nu\sigma + \partial_\mu\sigma\,\partial_\nu\varphi - \eta_{\mu\nu}\left(\frac{\sigma^2}{2}+\partial_{\gamma}\varphi\partial^{\gamma}\sigma\right)
\ee
is not invariant.  One would have to construct gauge-invariant operators from $\Box\varphi$ and insist that they alone are physically meaningful.
However, this is an artificial additional requirement that is not forced upon us {\it a priori}, and moreover it leads to a physically uninteresting
theory.

In the textbook Ref.\ \cite{Bogolyubov:1990kw}, the $(\Box\varphi)^2$ theory was meant to be a model for QED with covariant gauge fixing term (see section 10.1\,B). Then the full action, including the gauge fixing term $\sigma^2$, where $\sigma = \partial_\mu A^\mu$ (called $\Lambda$ in Ref.\ \cite{Bogolyubov:1990kw}) is invariant under gauge
transformations $\alpha$ that are harmonic.
This is an old pre-BRST\footnote{Becchi-Rouet-Stora-Tutyin \cite{Becchi:1975nq,Tyutin:1975qk}} approach where the gauge-fixed Lagrangian is treated as fundamental, and
only harmonic gauge transformations are considered to be symmetries.
In this approach, the QED stress tensor contains a term proportional to $\sigma$, but it does not contribute to
physical correlators since $\sigma$ commutes with all gauge-invariant operators and its vacuum expectation value is
zero [by fiat, (10.94)], {\it e.g.,} $\sigma \to 0$ in all observables, corresponding to the choice of gauge $\partial_\mu A^\mu=0$.

One can see that the theory only becomes trivial, in the sense of having no dynamical degrees of freedom,  when the extra condition $\sigma = 0$ is imposed.  Consider general plane-wave solutions,
\bea
%\label{EOMs}
    \sigma &=& a_k e^{-ik\cdot x} + a^*_k e^{i k\cdot x}\,;\\
    \varphi &=& (a_k\, \bar k\!\cdot\!x +b_k\, f(k\!\cdot\! x)) e^{-ik\cdot x}
    +{\rm h.c.}\,,\nn
\eea
where $k^\mu$ and $\bar k^\mu$ are null vectors ($k^2=\bar k^2 = 0$) with $\bar k^\mu = i(k^0,-\vec k)/(2|\vec k|^2)$, so that $i\bar k\cdot k = 1$.
These satisfy the equations of motion (\ref{EOMS}), since the combination
$fe^{-ik\cdot x}$ multiplied by $b_k$ is harmonic.   Without additional assumptions, the independent coefficients $a_k$ and $b_k$ correspond to two scalar degrees of freedom.  
If we treat $\varphi\to\varphi + \alpha(x)$ as a gauge symmetry, the $b_k$ term can be set to zero, but the
remaining $a_k$ terms can only be eliminated by declaring they are zero. {In other words, the above choice of gauge does not fix all of the degrees of freedom \cite{Hell:2023mph}.}
This is clearly not the situation that should be envisioned by Refs.\ \cite{Boyle:2021jaz,Turok:2023amx,Boyle:2025bxf}, which want the new scalars to be dynamical degrees of freedom, that contribute to the vacuum energy density and to density perturbations in the early Universe.

In Refs.\ \cite{Turok:2023amx,Boyle:2025bxf}, another work which was cited in connection to the supposed gauge symmetry was
Ref.\ \cite{Rivelles:2003jd}.  This paper shows that certain higher-derivative scalar theories can be made trivial by supplementing them with a ghost field $c$ that enables a BRST symmetry.  In particular, the Lagrangian
\be
    {\cal L} = \sfrac12\Box\varphi\Box\varphi - \bar c\Box c
\ee
has such a symmetry, which renders it trivial, in the sense that the only state it possesses is the ground state.  However, Refs.\ \cite{Turok:2023amx,Boyle:2025bxf} do not include the ghost field, which would entail a different theory than the one they purport to study.  Moreover, if they did so, the resulting theory would possess no fluctuations, and would not have any effect on the vacuum energy nor on cosmological perturbations.

In summary, this theory has no gauge symmetry, unless one adds to it an anticommuting ghost field, or by fiat declares that $\sigma$ is not observable and should be set to zero, which Ref.\ \cite{Bogolyubov:1990kw} imposed in order to turn it into a model of partially gauge-fixed QED.  Intrinsically both $\varphi$ and $\sigma$ are propagating degrees of freedom.

\section{The cosmological constant}

In Ref.\ \cite{Boyle:2021jaz}, the one-loop contribution of $\varphi$ to the vacuum energy was computed using the Euclidean Green's function, finding that $\varphi$ contributes the same as two normal massless scalars to the cosmological constant $\Lambda$.  By counting degrees of freedom, one can then cancel the ultraviolet contributions of the Standard Model particles to $\Lambda$, excluding those of the Higgs boson, by adding 36 $\varphi$ fields.  (It was hypothesized that the Higgs could be a bound state of standard model fermions, and would therefore not make a separate contribution.)  It is unclear how the results of this derivation would carry over to Minkowski space.

We again use the two-field formulation of Eq.\ (\ref{Stwo-derivative}) restricted to Minkowski space, in which computations simplify.  One can readily show that the propagator in the $(\varphi,\,\sigma)$ field space has the matrix form 
\bea
    G_{ij}(x,x') &=& i\int {d^{\,4}k\over (2\pi)^4}\left(\begin{array}{cc} {1\over k^4}& {1\over -k^2- is_1\epsilon}\\
        {1\over -k^2-is_2\epsilon} & 0 \end{array}\right)\nn\\
        &\times& e^{i k\cdot(x-x')}\,,
\label{propagator}
\eea 

where the the factors $s_i=\pm 1$ allow for general sign choices for the $i\epsilon$ prescriptions of the two fields. 
Notice that $\varphi$ and $\sigma$ have different dimensions, leading to 
different powers of $k^2$ amongst the matrix elements.  The $1/k^4$ term (shorthand for the product of the two off-diagonal propagators) does not contribute to the vacuum energy.
One can compute the expectation value of the Hamiltonian (\ref{Hameq})
from the propagator,
\bea
    \langle{\cal H}\rangle &=& \sfrac12\lim_{x\to x'}\Big[\partial_t\partial_{t'} +
    \vec\nabla_x\!\cdot\!\vec\nabla_{x'}\Big](G_{12}+G_{21})\nn\\
        &=& \sfrac12\int {d^{\,3}k\over (2\pi)^3} |\vec k|(s_1 + s_2)\,.
        \label{vac-energy}
\eea
If $s_1 = s_2 = +1$, this gives the result of Ref.\ \cite{Boyle:2021jaz},
namely twice the contribution of a normal scalar field.  However,
one of the fields is still a ghost, and choosing the sign of $\epsilon$ to give it positive energy endows it with negative
probability, causing violation of 
 unitarity in scattering amplitudes, as we will show in section \ref{unitarity}.

It is enlightening to repeat the calculation using field redefinitions that convert the Lagrangian to one involving conventionally normalized scalar fields with mass dimension one:
\be
\label{2ndparam}
    \varphi = {1\over 2\,M}(\eta + \psi),\quad
    \sigma = {M}(\eta-\psi)\,,
\ee
where $M$ is an arbitrary mass scale.  Then the Lagrangian becomes
\be
    {\cal L} = \sfrac12\Big( -(\partial\eta)^2 + (\partial\psi)^2 - M^2(\eta-\psi)^2\Big)\,.
\ee
In the $(\eta,\,\psi)$ basis, the matrix propagator in momentum space is 
\bea
    G_{ij} &=& {\rm diag}\left({i\over -k^2-is_1\epsilon}, {-i\over -k^2-i s_2\epsilon}\right) \nn\\ &+&
    i\,{M^2\over k^4}\left({ 1\atop 1}\,{1\atop 1}\right)\,,
\eea
and the Hamiltonian density is
\be
    {\cal H} = \sfrac12\left(\dot\eta^2 + |\nabla\eta|^2 - \dot\psi^2 -|\nabla\psi|^2   + M^2(\eta-\psi)^2\right)\,.
\ee
Because of the wrong sign for the $G_{22}$ element, the contribution of 
$\psi$ to the expectation value $\langle {\cal H}\rangle$ is the same as that of $\eta$ when $s_1=s_2$,
the sign in the propagator canceling the sign in ${\cal H}$, despite that $\psi$ is a ghost while $\eta$ is a normal excitation.  The contributions going as $M^2$ must cancel since the choice of $M$ is arbitrary and cannot
affect any observable quantity.  In particular, results should be continuous in the limit $M\to 0$.  Hence this gives the same result as Eq.\ (\ref{vac-energy}).

Even if we were to accept the presence of the ghost field, there are several other deficiencies in this mechanism.  First, it approximates all the SM fields as being massless.  Therefore only the leading quartic divergences to the cosmological constant are cancelled.  The subleading ones that are quadratic in
the cutoff and in the masses are uncanceled, leaving a residual contribution that is still many orders of magnitude above the observed value.

\begin{figure*}[t]
\centerline{\includegraphics[width=1.5\columnwidth]{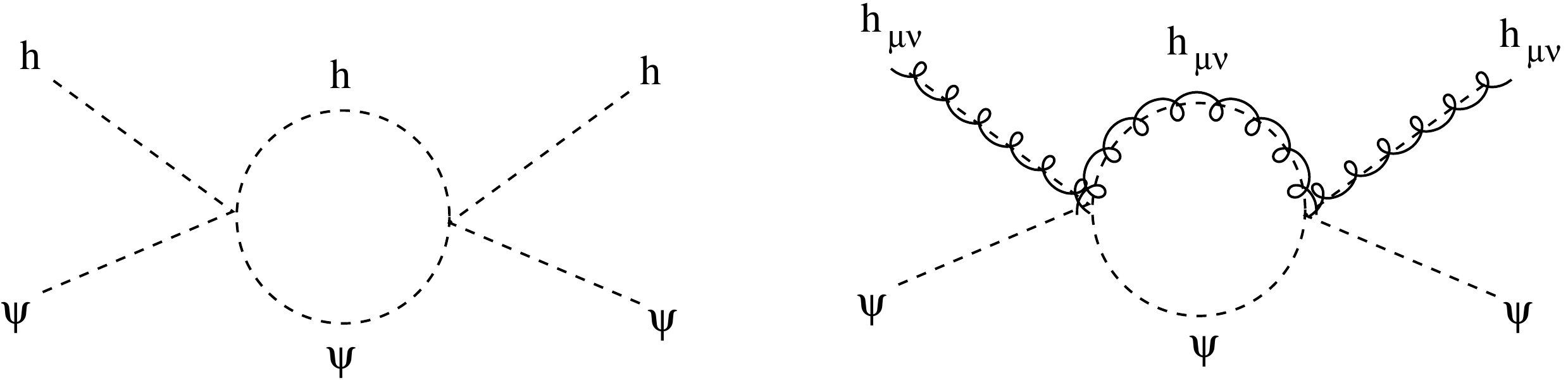}}
\caption{Left: hypothetical loop contribution to $\psi$-Higgs scattering, that violates unitarity because of the wrong-sign $\psi$ propagator.  Right: analogous unavoidable contribution to graviton-$\psi$ scattering, that similarly violates unitarity.}
\label{fig:diag}
\end{figure*}

Second, the enthusiasm of Ref.\ \cite{Boyle:2021jaz} is largely fueled by a numerical coincidence, that the addition of 36 dimension-zero scalars simultaneously cancels the Weyl anomaly and the cosmological constant, if one assumes that there are three right-handed neutrinos, {\it and} that the Higgs field is a composite state made of fields already within the fermionic sector of the SM.  The latter assumption is not supported by any theoretical studies that we are aware of; the known framework that comes closest is top quark condensation \cite{Cvetic:1997eb}, but this requires new degrees of freedom beyond the SM.  Moreover, the numerical coincidence only occurs if the possible linear couplings between $\varphi$ and curvature invariants, mentioned in the introduction in connection with Ref.\ \cite{Riegert:1984kt}, are omitted.  By including them, simultaneous cancellations of the Weyl anomaly and the quartically divergent contributions to the vacuum energy would be possible for any field content.

\section{Loss of unitarity}
\label{unitarity}

In this section, we will consider generic violations of unitarity that can occur in the higher-derivative theory as a consequence of the ghosts.
The field reparametrization (\ref{2ndparam}) makes it easier to appreciate these difficulties.  Let us consider a hypothetical coupling $\lambda h^2\psi^2$ of the Higgs field $h$ to the ghost field $\psi$.    Suppose that $s_2>0$ so that both $\psi$  carries positive energy.  Then the tree-level scattering $h\psi\to h\psi$ is allowed, but the $\psi$ propagator in the one-loop contribution has the wrong sign, and the imaginary part of this contribution to $h\psi\to h\psi$  will inherit this wrong sign, in violation of the optical theorem.  The Feynman diagram is shown in Fig.\ \ref{fig:diag} (left).

In the case where $s_2<0$, so that $\psi$ carries negative energy,
the scattering amplitudes for $h\psi\leftrightarrow h \psi$ are still kinematically allowed.  Unitarity is violated for the process $h\psi\to h\psi$ because of the wrong-sign propagator for $\psi$ in the loop, even though $\psi$ has a positive norm in the Hilbert space.  The imaginary part of the loop diagram does not depend upon $s_2$ in this case, since the Higgs has a finite width $\Gamma_h$.  When combining denominators using the Feynman parameter integral $x$, the imaginary part of the combined denominator becomes $i (x\Gamma_h + (1-x) s_2\epsilon)$, hence one can ignore $\epsilon$ in the loop.

We have illustrated the breakdown of unitarity in a model where $\varphi$ couples to another scalar field for simplicity, but even in the absence of such couplings, analogous interactions with gravitons are inevitable, and we expect the same problems to occur.

\section{Cosmological fluctuations and fifth forces}

Ref.\ \cite{Boyle:2021jaz} noted that the vacuum fluctuations of $\varphi$ take the form
\bea
    \langle\varphi^2\rangle &=& \lim_{x\to x'}G_{\varphi\varphi}(x,x') = \int {d^{\,4}k\over (2\pi)^4} {i\over k^4}\nn\\
    & \sim & \int {d^{\,3}k\over (2\pi)^3} {1\over k^3}
\eea
using Eq.\ (\ref{propagator}),
and doing the integral over $k^0$.  These are scale-invariant fluctuations, and so it is tempting to think that $\varphi$ could provide the cosmological seeds needed to get a nearly scale-invariant spectrum as observed in the cosmic microwave background (CMB), without the need to invoke inflation.  This idea was developed in Ref.\ \cite{Boyle:2025bxf}, where it was argued that $\varphi$ (in fact, 36 copies $\varphi_j$ of it) couples linearly to the energy density
$\rho_i$ of
each standard model field as a consequence of the trace anomaly.
\be
    {\cal L} \sim \sum_{ij} c_\chi \rho_i \varphi_j
    \label{phiint}\,,
\ee
where $c_\chi$ is dominated by the QCD contribution, $c_\chi \sim \alpha_s^2/(36\,\pi^2)$.  Then at high temperatures, the SM plasma sources fluctuations in $\varphi_i$
which induce a curvature perturbation, whose magnitude is argued to match that needed to explain the observed temperature fluctuations of the CMB .

Here we will not argue with the logic of these assertions; rather we point out that the interaction (\ref{phiint}) is ruled out by low-energy observables.  Extrapolating it to low energies, where $\alpha_s\sim 1$ and quarks become hadronized (coupling with similar strength to $\varphi_i$), the interaction with electrons and protons is of order 
\be
    {\cal L} \sim c_\chi\sum_i\varphi_i(m_e \bar e e + m_p \bar p p)
\ee
with $c_\chi\sim 1/(36\pi^2)$.  The exchange of $\varphi_i$ creates a fifth force between the electron and the proton, whose matrix element at low momentum transfer $\vec k^2$ goes as 
\be
    {\cal M} \sim 36\,c_\chi^2 m_e m_p (\bar u_e u_e) (\bar u_p u_p) {1\over {\vec k}^4}
\ee
in terms of the Dirac spinors, 
using the $G_{\varphi\varphi}$ element of the propagator (\ref{propagator}).  The corresponding potential energy is found by Fourier transforming ${\cal M}$, which by dimensional analysis corresponds to a linearly confining potential between the electron and the proton,
\be
    V({\vec x}) \sim 36\, c_\chi^2 m_e m_p |\vec x|\,.
    \label{linpot}
\ee
One can readily show that this is many orders of magnitude stronger than the Coulomb potential at distances of order the Bohr radius; hence the properties of atoms would be drastically changed by such a force.

\section{Conclusions}
The appearance of zero-dimension scalar fields in conformal supergravity, a theory with interesting properties that might have recommended it as a consistent unification with gravity, motivates their further study.  In this paper we have made a thorough analysis of the conformal $\varphi\Delta_4\varphi$ theory, both in Minkowski space and in gravitational backgrounds, either considering the background to be fixed or fluctuating.

As one would expect, the theory, being fourth order in derivatives, has a ghost in any of these settings. In the general case, the theory propagates two scalar degrees of freedom, out of which at least one is a ghost mode.  When the system is coupled to gravity, it can happen that both scalars become ghosts.  This is the case when $\varphi$ has a classical background that dominates the expansion, which we showed simulates a radiation-dominated Universe.  If $\varphi$ partially dominates the expansion, the situation is complicated, but always contains at least one ghost.  In the limit where $\varphi$ becomes a spectator field, the system has similar behavior to Minkowski space, with one ghost and one non-ghost field.

At the quantum level, we have shown that the ghosts indeed lead to violation of unitarity of scattering amplitudes, and that the cancellation of the cosmological constant is incomplete at best. Moreover, the couplings of $\varphi$ to the SM, needed to get the observed spectrum of CMB fluctuations, is ruled out by the drastic effect that $\varphi$ exchange would have on atomic structure.  The dimension zero scalar theory thus creates serious problems of its own, without offering any viable solutions to existing ones.

\bigskip
{\bf Acknowledgments.}  We thank Simon Caron-Huot for valuable contributions at the start of this work and for comments on the manuscript; also A.\ Tseytlin for helpful comments and pertinent information.  The work of JC is supported by the Natural Sciences and Engineering Research Council (NSERC) of Canada.  JC thanks the Niels Bohr International Academy for its hospitality during the inception of this study. A.H. would like to thank Robert Brandenberger for useful discussions, and the McGill University for hospitality, where part of this work was carried out. The work of A.H. was in part supported by the KAKENHI No.25H00403, and is supported by the World Premier International Research Center Initiative (WPI), MEXT, Japan. 

\begin{appendix}

\section{Coefficients for Section \ref{FLRW}}
Here we provide the full expressions for the coefficients for the scalar modes corresponding to the equation (\ref{scalar-pert}). They are given by:\onecolumngrid  \begin{equation}
    c_1(t,k^2)=-\frac{24 \dot{\varphi}^{2} a^{5}}{\left(32 k^2+24 a \ddot{a}+3 \dot{a}^{2}\right) \dot{\varphi}^{2}+12 \dot{\varphi} \dot{a} \ddot{\varphi} a+36 {\ddot{\varphi}}^{2} a^{2}},
\end{equation}
and
\begin{equation}
   \begin{split}
        \gamma(t,k^2)c_2(t,k^2)&= -\left(\left(-\frac{15 \dddot{a} \dot{a}}{16}+\frac{17 \ddot{a}^{2}}{8}\right) a^{2}+\frac{43 \left(k^{2}+\frac{93 \dot{a}^{2}}{172}\right) \ddot{a} a}{12}+k^{4}+\frac{313 k^{2} \dot{a}^{2}}{96}+\frac{17 \dot{a}^{4}}{32}\right) \dot{\varphi}^{4}\\&+\frac{11 a \ddot{\varphi} \left(\dot{a} k^{2}+\frac{5}{11} \dddot{a} a^{2}+\frac{37}{11} \dot{a} \ddot{a} a +\frac{24}{11} \dot{a}^{3}\right) \dot{\varphi}^{3}}{8}+\frac{13 a^{2} \left(\frac{16 a \ddot{a}}{13}+k^{2}+\frac{185 \dot{a}^{2}}{52}\right) \ddot{\varphi}^{2} \dot{\varphi}^{2}}{8}\\&+\frac{11 \ddot{\varphi}^{3} a^{3} \dot{a} \dot{\varphi}}{16}-\frac{3 a^{4} \ddot{\varphi}^{4}}{8},
   \end{split}
\end{equation}
with
\begin{equation}
    \gamma(t,k^2)=\frac{k^2\left(36 \ddot{\varphi}^{2} a^{2}+12 \dot{a} \dot{\varphi} \ddot{\varphi} a +24 \ddot{a} \dot{\varphi}^{2} a +32 \dot{\varphi}^{2} k^{2}+3 \dot{a}^{2} \dot{\varphi}^{2}\right)^{2}}{1152  a^{3}}
\end{equation}
\twocolumngrid
In the high-$k$ limit, the above two expressions simplify, giving
\begin{equation}
   c_1(t,k^2)\sim-\frac{3a^5}{4k^2}, 
\end{equation}
and
\begin{equation}
    c_2(t,k^2)\sim-\frac{9a^3}{8k^2}. 
\end{equation}

To find the speed of propagation, we find the corresponding equation of motion,
and assume the plane-wave solution $\sigma\sim e^{-i\omega t}$. By further setting $\omega=\frac{k}{a}\sqrt{c_s}$, where $c_s$ is the speed of propagation, we arrive to the following value in the high-k limit:  
\begin{equation}
    c_s^2=\frac{3}{2}
\end{equation}

For the special case of Minkowski space $a=1$ with the exact solution $\dot\phi=1$, we find
\be
    c_1 = -{3\over 4},\quad c_2 = -{9\over 8}\,,
\ee
giving a propagation speed of $\sqrt{3/2}$ and a ghost-like kinetic term.

\subsection{Tensor fluctuations in the $\Delta_4$ theory without the Einstein term}

The Lagrangian density for the tensor perturbations corresponding to the action (\ref{D4sc}) is given by
\begin{equation}
    \mathcal{L}_T=-a\dot{\varphi}^2\left(\frac{1}{3}a^2\dot{h}_{ij}^T\dot{h}_{ij}^T+\frac{1}{6}h_{ij,k}^Th_{ij,k}^T\right)\,.
    \label{tensor-fluc}
\end{equation}
The tensor fluctuation is thus a ghost at all wavenumbers, and it is also tachyonic because of the relative $+$ sign between the two terms.
The squared propagation speed $v_s^2 = -1/2$ is subluminal in
magnitude.
The kinetic term vanishes as $\dot{\varphi}\to 0$, signaling a possible strong-coupling obstruction when canonically normalizing $h^T_{ij}\to h^T_{ij}/|\dot\varphi|$ and accounting for the gravitational couplings of $h^T_{ij}$ at cubic and higher order.  This is analogous to the van Dam-Veltman-Zakharov (vDVZ) 
discontinuity \cite{vanDam:1970vg,Zakharov:1970cc} that separates massive and massless gravity theories {at linear order in the perturbations}. {Although this discontinuity might be resolved through the Vainshtein mechanism similarly to the massive gravity, or a similar mechanism in the massive Yang-Mills theory \cite{Vainshtein:1972sx, Hell:2021oea}, here we will not further pursue this aspect of the theory in the light of the sicknesses which afflict it already on the level of the scalar perturbations. } 

\end{appendix}

\bibliographystyle{utphys}
\bibliography{paper}

\end{document}